\documentclass[conference]{IEEEtran}
\IEEEoverridecommandlockouts
\usepackage{cite}
\usepackage{algorithmic}
\usepackage{algorithm} 
\usepackage{amsmath,amssymb,amsfonts}
\usepackage{graphicx}
\usepackage[caption=false]{subfig}
\usepackage{textcomp}
\usepackage{enumerate}
\usepackage[nolist]{acronym}
\usepackage[table,xcdraw]{xcolor}
\usepackage{breqn}
\usepackage[dvipsnames]{xcolor}
\usepackage{multicol}
\usepackage{subcaption}
\usepackage[export]{adjustbox}
\usepackage{capt-of}
\usepackage{caption}
\usepackage{booktabs}
\usepackage{float}
\usepackage{multirow}
\usepackage{cleveref}
\usepackage{enumitem}
\usepackage[export]{adjustbox}
\newcommand{\squeezeup}{\vspace{-2mm}}
\newcommand{\mini}{\vspace{-1mm}}

\usepackage[hyphens]{url}
\usepackage{url}

\usepackage[absolute,showboxes]{textpos}

\newcommand{\third}{\texttt{PyRBD3}}
\newcommand{\fast}{\texttt{Fast-MCS}}
  
\TPMargin{5pt}

\newcommand{\copyrightstatement}{
    \begin{textblock}{0.84}(0.08,0.01)    
         \noindent
         \footnotesize
         \copyright This work has been submitted to the IEEE for possible publication. Copyright may be transferred without notice, after which this version may no longer be accessible
    \end{textblock}
}

\def\BibTeX{{\rm B\kern-.05em{\sc i\kern-.025em b}\kern-.08em
    T\kern-.1667em\lower.7ex\hbox{E}\kern-.125emX}}

\begin{document}

\bstctlcite{IEEEexample:BSTcontrol}

\title{\texttt{Fast-MCS}: A Scalable Open-Source Tool to Find Minimal Cut Sets\\
\thanks{This work has been funded by the Federal Ministry of Research, Technology, and Space in Germany (BMFTR) under the project SUSTAINET-ADVANCE (grant ID: 16KIS2279 and 16KIS2283) and the Bavarian Ministry of Economic Affairs, Regional Development, and Energy under the project `6G Future Lab Bavaria'.}}

\author{\IEEEauthorblockN{Shakthivelu Janardhanan\textsuperscript{*}, Yaxuan Chen\textsuperscript{*}, Wolfgang Kellerer\textsuperscript{*}, and Carmen Mas-Machuca\textsuperscript{*\textdagger}}
\IEEEauthorblockA{\textsuperscript{*}Chair of Communication Networks (LKN), Technical University of Munich (TUM), Germany}
\IEEEauthorblockA{\textsuperscript{\textdagger}Chair of Communication Networks, University of the Bundeswehr Munich (UniBW), Germany \\
\{shakthivelu.janardhanan, yaxuan.chen, wolfgang.kellerer\}@tum.de, cmas@unibw.de}
\squeezeup \squeezeup \squeezeup \squeezeup}

\maketitle
\copyrightstatement
\IEEEpubidadjcol

\begin{abstract}
A network is represented as a graph consisting of nodes and edges. A cut set for a source-destination pair in a network is a set of elements that, when failed, cause the source-destination pair to lose connectivity. A \ac{MCS} is a cut set that cannot be further reduced while maintaining its status as a cut set. \acp{MCS} are crucial in identifying the critical elements in the network that have the most significant impact on failure. This work introduces \fast, an open-source, scalable tool for evaluating \acp{MCS} in large, complex networks. Additionally, we compare the computation time of \fast~with the state-of-the-art.
\end{abstract}


\section{Introduction}
\label{chap:intro}

Modern networks, such as communication networks, electrical power grids, transportation, water supply, gas pipelines, etc., can be represented as graphs, consisting of nodes and edges. These networks invariably require high levels of connectivity consistently. A network operator needs to know the most critical elements of the network so they can be better protected in the event of failures. 

\begin{figure}[]
    \centering
    \includegraphics[width=0.8\linewidth]{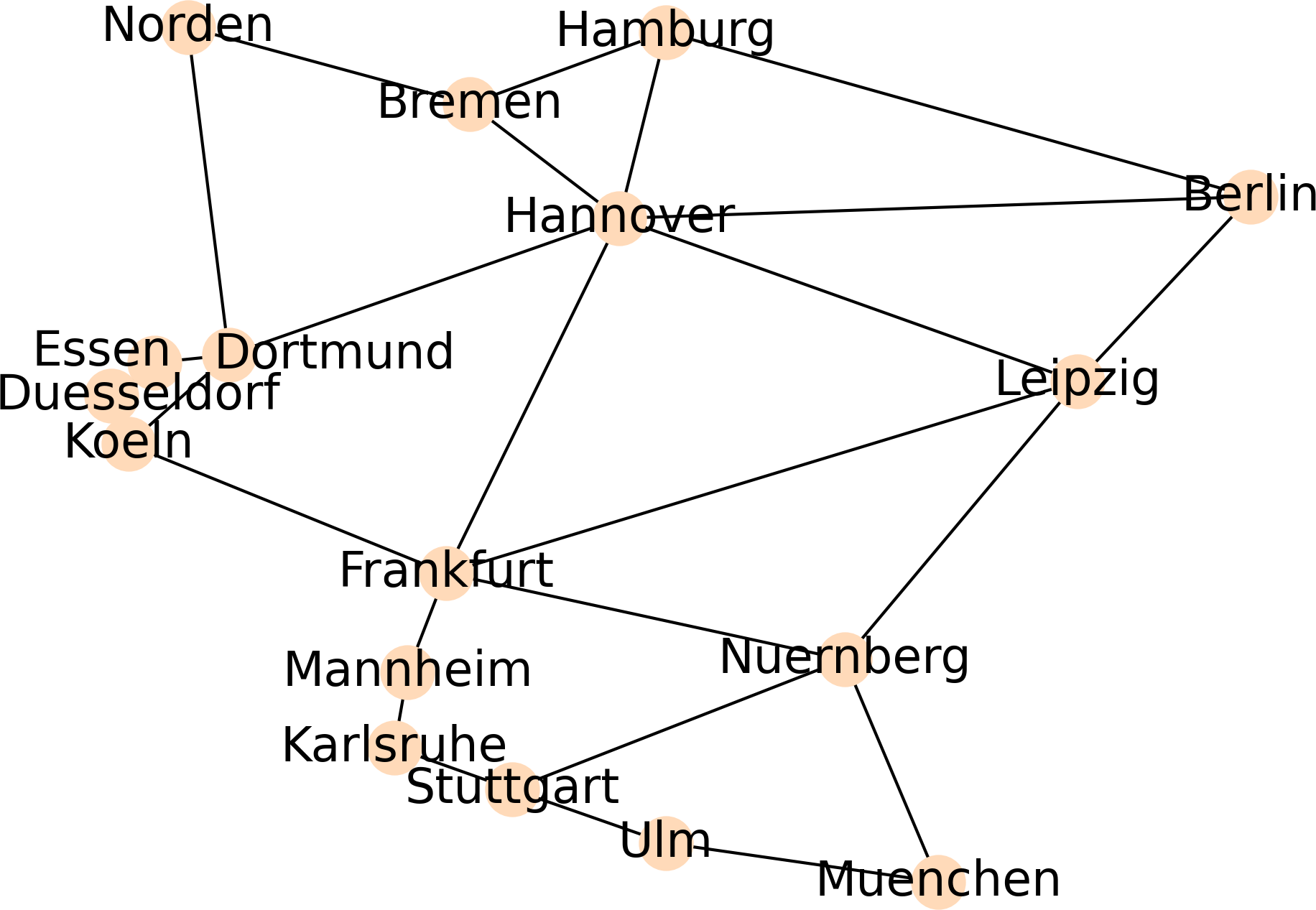}
    \caption{Germany\_17 topology~\cite{sndlib}}
    \label{fig:topo_Germany_17}
\end{figure}

The connectivity between any source and destination relies on how they are connected in the topology. When a set of elements in the network fails, disconnecting the source-destination pair, that set is called a cut set. A \ac{MCS} is also a cut set that cannot be further reduced~\cite{mcs}. For example, consider the Germany\_17 topology~\cite{sndlib} shown in Fig.~\ref{fig:topo_Germany_17}. Let us assume some traffic between the source Ulm and the destination Hamburg. For this source-destination pair, Ulm-Hamburg, (Frankfurt, Leipzig, Hannover) is a cut set because the failure of these nodes causes Ulm and Hamburg to be disconnected. However, the set can be further reduced to (Frankfurt, Leipzig) and still maintain its status as a cut set. Hence, (Frankfurt, Leipzig) is a \acf{MCS}. Note that a single source-destination pair can have several \acp{MCS} of varying sizes. For example, (Dortmund, Hannover, Leipzig) is another \ac{MCS} of Ulm-Hamburg with size three. Moreover, a node can be part of several \acp{MCS}. In this example, Leipzig is a part of both the \acp{MCS} discussed. 

\acp{MCS} are a popular tool used in modern networks for various purposes across various fields, including aerospace, nuclear power, transportation, electrical infrastructure, communication networks, etc. For example, \acp{MCS} are used in Fault Tree Analysis~\cite{ft_mcs,faulttree_analaysis_1,fiabilipy, ft2, relyence}, \ac{RBD} and availability evaluation~\cite{mine_pyrbdplusplus, mine_pyrbd3, 7}, sovereignty metric evaluation~\cite{mine6_ICTON-demo, mine_commag}, resilience evaluation~\cite{mine_GUI, mine_PyRobust}, etc. Note that the network type does not change the concepts discussed in this work. Our \textit{Fast-MCS} solution is applicable to any network or system that can be represented as a graph consisting of components as nodes and their interconnections as edges. However, this manuscript focuses on communication networks as the primary example.

Several nodes and edges exist in large communication networks. For a source and destination in the network, finding the \acp{MCS}, i.e., finding the smallest combinations of nodes whose failure can cause loss of connection between that source and destination, gets computationally expensive as the size of the topology increases. This problem is noted to be NP-hard~\cite{yeh}. The aforementioned tools and evaluations depend on finding the \acp{MCS} first. Therefore, it is essential to find a fast algorithm to evaluate \acp{MCS} based on the topology information. 

Hence, we present the following contributions in this work. 
\begin{enumerate}
\item Sec.~\ref{chap:related} investigates methods to evaluate \acp{MCS}.
\item Sec.~\ref{chap:method} discusses our novel method to evaluate find \acp{MCS}, and its implementation as an open-source tool called \fast.
\item Sec.~\ref{chap:results} shows the performance of \third, and compares it against the state-of-the-art.
\end{enumerate}

\section{Background and Related Work}
\label{chap:related}

The most intuitive method to find \acp{MCS} is a combinatorial searching algorithm, i.e., checking if each combination of nodes in the network could be a \ac{MCS}. \cite{mine_pyrbd,mine_pyrbdplusplus,alghanim} propose extensions to the classical combinatorial searching algorithm to speed up the process. The authors in~\cite{kvassay_analysis_2016} extend this approach using Direct Partial Boolean Derivatives (DPBDs). They study the impact of changing a component's state on the overall system availability, with positive and negative effects on path and cut sets, respectively.
However, the aforementioned approaches do not scale well for large topologies because the number of possible node combinations scales as $\sum\limits_{i=1}^{|V|-1}\binom{|V|}{i}$.

Other previous works also consider evaluating \acp{MCS}. Most of these works discuss the evaluation of \acp{MCS} based on \acp{MPS}.
A path set between a source and a destination is the complement of a cut set; i.e., a path set is a set of nodes whose availability connects the source and destination. A \ac{MPS} is also a path set, but it cannot be further reduced. For example, let us assume some traffic between the source Ulm and the destination Hamburg, in the Germany\_17 topology in Fig.~\ref{fig:topo_Germany_17}. For this source-destination pair, \{Ulm, Muenchen, Nuernberg, Leipzig, Berlin, Hannover, Hamburg\} is a path set because the availability of these nodes ensures a successful path between the source and the destination. However, this path set is not minimal because it can be further reduced to \{Ulm, Muenchen, Nuernberg, Leipzig, Berlin, Hamburg\}. This new set cannot be reduced further, and hence is a \ac{MPS}. Similar to \acp{MCS}, a single source-destination pair can have several \acp{MPS}. For example, \{Ulm, Stuttgart, Nuernberg, Leipzig, Berlin, Hamburg\} is another \ac{MPS}. Additionally, a single node can be a part of multiple \acp{MPS}. 

\acp{MPS} and \acp{MCS} are the prime implicants of monotone Boolean functions~\cite{43} for the success and the failure of a system, respectively. They are also referred to as the minimal satisfying and maximal non-satisfying vectors of a Boolean function~\cite{46}.
Since a \ac{MPS} is the logical opposite of a \ac{MCS}, finding the logical complement of the \ac{MPS} yields the \ac{MCS}. 

In~\cite{heidtmann}, the authors introduced the INMIN algorithm for inverting path sets and cut sets. However, the approach relies on exhaustive power-set enumeration and does not scale to large systems. Furthermore, the INMIN algorithm only evaluates the cut sets. It does not evaluate \acp{MCS}. This would require further computation. The authors in~\cite{rai, locks} introduced a recursive inversion technique based on De Morgan’s laws and Boolean absorption to convert \acp{MPS} to \acp{MCS}. Other works, such as~\cite{prasad, lamalem}, use similar approaches to find \acp{MCS}. 

In~\cite{rushdi2021derivation}, the authors present their approach based on De Morgan's rules and Boole-Shannon expansion to evaluate \acp{MCS} from \ac{MPS}. Their primary advantage is the ability to split the expression into smaller partitions for faster calculation and recursive absorption. Therefore, the computation required to obtain the minimum-sum-of-products expression of the \acp{MCS}. Their approach is efficient for smaller networks. However, it does not scale for larger networks. This method is further explained in Sec.~\ref{sec:shannon}. 

To overcome the issues identified in previous works, we introduce the following improvements.
\begin{enumerate}
    \item Our method relies on logically complementing the \acp{MPS} by using De Morgan's laws and selects a single common element per step to enable scalability across large topologies.
    \item Our method uses a faster, more memory-efficient set-theoretic approach to complement \acp{MPS}, rather than the long Boolean expressions used in~\cite{rushdi2021derivation}. 
    \item We implement our method as an open-source tool called \fast~\cite{mine_git_fastmcs}.
    \item We compare the performance of \fast~against the combinatorial method~\cite{mine_pyrbd,mine_pyrbdplusplus,alghanim} and the Boole-Shannon method~\cite{rushdi2021derivation}.
\end{enumerate}
The implementation of the combinatorial method~\cite{mine_pyrbd,mine_pyrbdplusplus,alghanim} and the Boole-Shannon expansion~\cite{rushdi2021derivation} are explained in Sections~\ref{sec:combi} and~\ref{sec:shannon}, respectively. Note that this manuscript considers only nodes to facilitate easier explanation. The concepts of \acp{MCS} and \acp{MPS} apply to links as well as edges. Furthermore, our tool \fast~can include links in the \acp{MCS} evaluation, if required.

\begin{figure*}[h]
    \centering
    \includegraphics[width=0.7\linewidth]{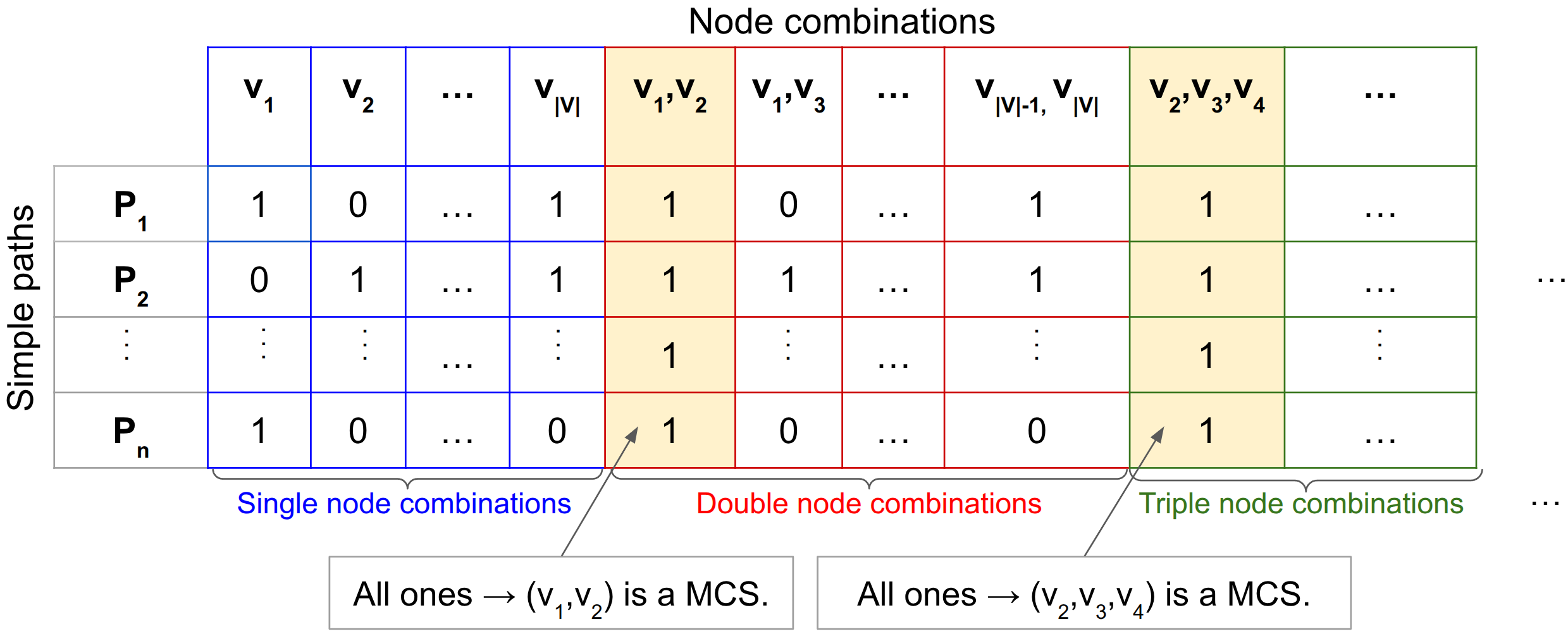}
    \caption{Combinatorial procedure to evaluate \acfp{MCS}.}
    \label{fig:combi} \squeezeup
\end{figure*}
\section{Methodology}
\label{chap:method}

This section discusses the implementation of the combinatorial method~\cite{mine_pyrbd, mine_pyrbdplusplus, alghanim}, the Boole-Shannon expansion~\cite{rushdi2021derivation} technique, and our \fast. For the Boole-Shannon expansion~\cite{rushdi2021derivation} and \fast, we use the same algorithm to find \acp{MPS}.

\subsection{Combinatorial method to find \acp{MCS}}
\label{sec:combi}
The combinatorial method~\cite{mine_pyrbd, mine_pyrbdplusplus, alghanim} to find the \acp{MCS} for a source-destination pair is explained in Fig.~\ref{fig:combi}. First, all the simple paths from the source to the destination are noted down as rows. Second, all nodes except the source and destination are treated as potential \acp{MCS} and listed as columns (in blue). The table is populated as follows.
\begin{enumerate}
\item If node $v_i$ is used in path $P_i$, then the corresponding value is one.
\item If node $v_i$ is not used in path $P_i$, then the corresponding value is zero.
\end{enumerate}
After populating all the blue columns, if any column has all ones, then the corresponding node is present in all the simple paths from the source to the destination. Therefore, if that node fails, connectivity between the source and the destination will be lost. Hence, this node is a \ac{MCS}. 

Next, all possible two-node combinations are taken as columns (in red). The supersets of previously identified \acp{MCS} are removed because these supersets can only be cutsets and not \acp{MCS}. Then, the columns can be populated by using the logical `OR' operation between the respective columns of the nodes in the combination. For example, the column for $v_i, v_j$ is obtained by the logical `OR' operation between the values in the columns for $v_i$ and $v_j$. After populating the red columns, the test for \acp{MCS} is performed again by checking the existence of columns with all ones. 

This procedure of populating columns for different node combinations and checking for \acp{MCS} is repeated until no more columns are possible. A collection of all the all-ones columns provides all the \acp{MCS} for that source-destination pair. For example, in Fig.~\ref{fig:combi}, $(v_1, v_2)$ and $(v_2,v_3,v_4)$ are \acp{MCS}.

\subsection{Finding \acfp{MPS}}
\label{sec:findmps}
We find the \acp{MPS} based on an extension of the \ac{DFS} algorithm seen in Alg.~\ref{algo:1}~\cite{mine_pyrbd3}.
First, the branching starts at the source node. The newly explored node is added to the list of visited nodes in that branch. 
At every step of the branching procedure, two conditions are checked:
\begin{enumerate}
    \item if the newly explored node is the destination, or 
    \item if the newly explored node has already been visited in that branch.
\end{enumerate} 
If at least one of the aforementioned conditions is true, the branch is terminated. If both conditions are not met, the branching procedure is continued. 

Consider Ulm-Hamburg as the example source-destination pair. In the path set \{Ulm, Muenchen, Nuernberg, Leipzig, Berlin, Hannover, Hamburg\}, Hamburg is reached from Berlin via Hannover. However, Hamburg can be reached directly from Berlin. Therefore, the proposed path set is not minimal. In other words, when a node that can be reached from one of the previously visited nodes through a shorter path is visited through a longer path, that path is no longer minimal. 

Hence, the \ac{DFS} algorithm must be modified to terminate a branch if a new node is being visited that can already be visited from some other node in the visited list directly. This modification to the \ac{DFS} algorithm, shown in line 10 of Alg.~\ref{algo:1}, speeds up the search for \acp{MPS}. The branching procedure continues until all branches have been explored and terminated.

\begin{algorithm}
\caption{Find \acfp{MPS} based on the modified \acf{DFS} algorithm.}
\label{algo:1}
\begin{algorithmic}[1]
\REQUIRE Graph $G = (V,E)$, source $src$, destination $dst$.
\ENSURE $\mathcal{P}$ is the list of all \acp{MPS} from $src$ to $dst$. $\mathcal{P} \gets \emptyset$
\STATE \textbf{Call} DFS($s$, $\emptyset$)
\vspace{0.4cm}

\STATE \textbf{function} modified\_DFS($current\_n$, $path\_p$)
\STATE \quad Add $current\_n$ to $path\_p$.
\STATE \quad \textbf{If} $current\_n = dst$, \textbf{then}
\STATE \quad \quad Store $path\_p$ in $\mathcal{P}$.
\STATE \quad \quad Remove $current\_n$ from $path\_p$. 
\STATE \quad \quad \textbf{return}
\STATE \quad \textbf{end if}
\STATE \quad \textbf{For} each neighbor $neighbor\_n$ of $current\_n$ in $G$, \textbf{do}
\STATE \quad \quad \textbf{If} $neighbor\_n \notin path\_p$ \textbf{and} $neighbor\_n$ cannot be reached directly from another node in $path\_p$, \textbf{then}
\STATE \quad \quad \quad \textbf{call} modified\_DFS($neighbor\_n$, $path\_p$)
\STATE \quad \quad \textbf{end if}
\STATE \quad \textbf{end for}
\STATE \quad Remove $current\_n$ from $path\_p$.
\STATE \quad \textbf{return}
\end{algorithmic} 
\end{algorithm}


\begin{figure}[]
    \centering
    \includegraphics[width=0.8\linewidth]{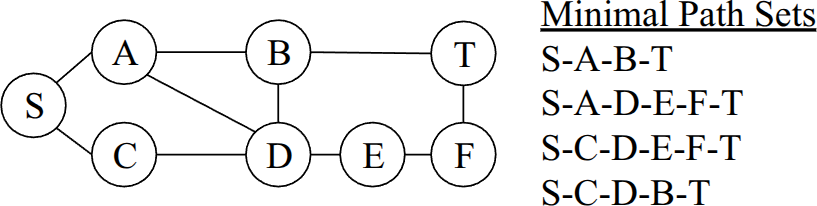}
    \caption{Example topology, source-destination pair, and its corresponding \acfp{MPS}.}
    \label{fig:topo}
\end{figure}

\begin{figure*}
    \centering
    \includegraphics[width=0.65\linewidth]{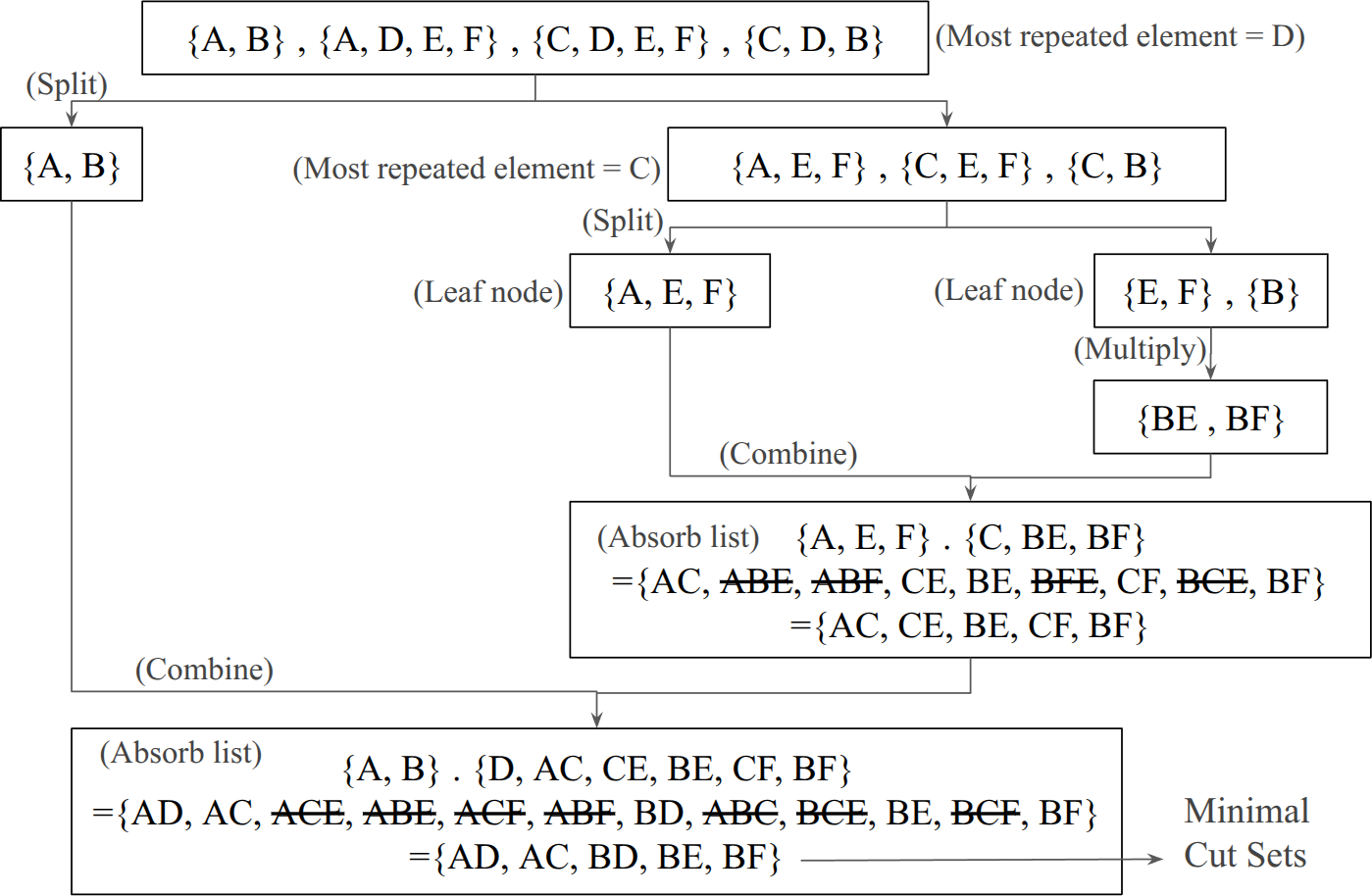}
    \caption{\fast's algorithm illustrated with the example source-destination pair, $S-T$, in Fig.~\ref{fig:topo}.}
    \label{fig:fastexampleexplained} 
\end{figure*}

\subsection{Finding \acfp{MCS} based on Boole-Shannon expansion} 
\label{sec:shannon}

In this method, the system's success, i.e., the source-destination pair's successful connectivity, is represented as its \acp{MPS}. Failure is represented by Boolean complementation~\cite{rushdi2021derivation}. 
A pivot element that appears most frequently is chosen to apply the Boole-Shannon expansion. Shannon expansion is applied as follows.
\begin{equation}
S=xS_{x=1}+\bar{x}S_{x=0},
\end{equation}
where $x$ is the pivot element. The $S_{x=1}$ and $S_{x=0}$ expressions are recursively split using the Boole-Shannon expansion, until there are no unique pivot elements.
The resulting final expression can be simplified and checked for absorption to obtain the \acp{MCS}. For example, in the first step, considering the example shown in Fig.~\ref{fig:topo}, for the source-destination pair S-T, the success function is,
\begin{equation}
S = AB + ADEF + CDEF + CDB.
\end{equation}
If $D$ is the pivot element, the next step splits $S$ as,
\begin{equation}
S_{D=1} = AB + AEF + CEF + CB,
\end{equation}
\begin{equation}
S_{D=0} = AB.
\end{equation}
Then the expressions are complemented and considered for further expansion. The final expression is reduced by Boolean absorption to obtain the \acp{MCS}.


\subsection{Finding \acfp{MCS} using \fast}
\label{sec:fast-mcs}


\begin{algorithm*}
\caption{\fast's \acf{MCS} computation using a Decision Tree}
\label{algo:fast}
\begin{multicols}{2}
\begin{algorithmic}[1]

\REQUIRE Set of minimal paths $\mathcal{P}$ from $S$ to $T$
\ENSURE Set of minimal cut sets $\mathcal{M}$

\STATE Remove $S$ and $T$ from each path in $\mathcal{P}$

\STATE $\mathcal{C}$ is the set that represents all \acfp{MPS}, where each \ac{MPS} $p$ is represented as a set. \\
$\mathcal{C} = \{ \text{set}(p) \mid p \in \mathcal{P} \}$ \COMMENT{CNF: minimal paths}

\STATE Begin Decision Tree with $\mathcal{C}$ as the root node $\mathcal{R}$. \\ \textbf{Call} $\mathcal{T}$ = \textit{Binary\_Decision\_Tree($\mathcal{R} = \mathcal{C}$).}

\STATE Evaluate the tree bottom-up. Start traversal from the root node $\mathcal{R}.$\\
\textbf{Call} $\mathcal{M}$ = \textit{Evaluate($\mathcal{R}$)}

\RETURN $\mathcal{M}$

\vspace{0.4cm}

\STATE \textbf{function} \textit{Binary\_Decision\_Tree($\mathcal{R}$)}
\STATE Initialize node $N$ with value $\mathcal{R}$.
\STATE Count frequency of each element in $\mathcal{R}$.
\STATE $x =$ most frequent element.
\IF{$x$ appears in more than one term in $\mathcal{R}$}
    \STATE \textbf{Call} $\mathcal{R}_{with}, \mathcal{R}_{without} = \textit{Split}(\mathcal{R}, x)$
    \STATE Create a new node in the Binary Decision Tree on the left of node $N$, \\ \textbf{Call} $N.left = \textit{Binary\_Decision\_Tree}(\mathcal{R}_{without})$
    \STATE Create a new node in the Binary Decision Tree on the right of node $N$, \\ \textbf{Call} $N.right = \textit{Binary\_Decision\_Tree}(\mathcal{R}_{with})$
    \STATE Store $x$ in $N$
\ELSE
    \STATE Node $N$ is a leaf node.
\ENDIF

\vspace{0.4cm}

\STATE \textbf{function} \textit{Split($\mathcal{R}$, x)}
\STATE $\mathcal{R}_{with} = \emptyset$
\STATE $\mathcal{R}_{without} = \emptyset$
\FOR{each term $\tau \in \mathcal{R}$}
    \IF{$x \in \tau$}
        \STATE $\mathcal{R}_{with} = \mathcal{R}_{with} \cup \{\tau \setminus \{x\}\}$
    \ELSE
        \STATE $\mathcal{R}_{without} = \mathcal{R}_{without} \cup \{\tau\}$
    \ENDIF
\ENDFOR
\RETURN $\mathcal{C}_{with}, \mathcal{C}_{without}$

\vspace{0.4cm}

\STATE \textbf{function} \textit{Evaluate($N$)}
\STATE Start evaluation at the root node $N$.
\IF{$N$ is a leaf}
    \RETURN \textit{EvaluateSelf}$(N)$
\ELSE
\STATE $\mathcal{L}eft = \textit{Evaluate}(N.left)$ if $N.left$ exists
\STATE $\mathcal{R}ight = \textit{Evaluate}(N.right)$ if $N.right$ exists
\STATE Combine $\mathcal{L}eft$ and $\mathcal{R}ight$ along with the most frequent element $x$ stored in node $N$ (from line 9) to get the updated value of current node $N$.\\
\STATE $N.\textit{evaluation} = \textit{Combine}(\mathcal{L}eft, \mathcal{R}ight, N.x)$
\ENDIF
\RETURN $N.\textit{evaluation}$

\vspace{0.4cm}

\STATE \textbf{function} \textit{Combine$(\mathcal{L}eft, \mathcal{R}ight, N.x)$}
\STATE $\mathcal{Q} = \emptyset$ (AbsorbList)
\IF{$x$ exists}
    \STATE Add $\{x\}$ to $\mathcal{Q}$
\ENDIF
\IF{$\mathcal{L}eft$ is empty}
    \STATE Add all sets in $\mathcal{R}ight$ to $\mathcal{Q}$
\ELSIF{$\mathcal{R}ight$ is empty}
    \STATE Add all sets in $\mathcal{L}eft$ to $\mathcal{Q}$
\ELSE
    \FOR{each $l \in \mathcal{L}eft$}
        \FOR{each $r \in \mathcal{R}ight$}
            \STATE Add $l \cup r$ as a new element to $\mathcal{Q}$.
        \ENDFOR
    \ENDFOR
\ENDIF
\RETURN $\mathcal{Q}$

\vspace{0.4cm}

\STATE \textbf{function} \textit{EvaluateSelf$(N)$}
\STATE Leaf node with value $\mathcal{X} = \{X_1, X_2, \dots, X_k\}$
\STATE $\mathcal{L}eafnode = \{\{e\} \mid e \in X_1\}$
\FOR{$i = 2$ to $k$}
    \STATE $\mathcal{L}eafnode = \textit{Multiply}(\mathcal{L}eafnode, \mathcal{X}_i)$
\ENDFOR
\IF{node has stored split element $x$}
    \STATE $\mathcal{L}eafnode = \mathcal{L}eafnode \cup \{\{x\}\}$
\ENDIF

\vspace{0.4cm}

\STATE \textbf{function} \textit{Multiply$(A, B)$}
\STATE $\mathcal{Z} \leftarrow \emptyset$. \STATE $\mathcal{Z}$ maintains minimality by removing supersets (AbsorbList).

\FOR{each set $a \in A$}
    \IF{$a \cap B = \emptyset$}
        \FOR{each $b \in B$}
            \STATE Add $a \cup \{b\}$ to $\mathcal{Z}$
        \ENDFOR
    \ELSE
        \STATE Add $a$ to $\mathcal{Z}$ \COMMENT{Absorb }
    \ENDIF
\ENDFOR

\end{algorithmic}
\end{multicols}
\end{algorithm*}


The methodology used in \fast~is shown in Alg.~\ref{algo:fast}. It is also explained using the example topology in Fig.~\ref{fig:topo}. Consider the source-destination pair $S-T$. The \acp{MPS} for this source-destination pair are first evaluated based on the method discussed in Sec.~\ref{sec:findmps}. For the example in Fig.~\ref{fig:topo}, $S-T$ pair has four \acp{MPS}: $S-A-B-T$, $S-A-D-E-F-T$, $S-C-D-E-F-T$, and $S-C-D-B-T$. After removing the source and destination nodes from the \acp{MPS}, the \acp{MPS} are represented as sets: $\{A,B\}$, $\{A,D,E,F\}$, $\{C,D,E,F\}$, and $\{C,D,B\}$. This is the first step shown in the illustrated example in Fig.~\ref{fig:fastexampleexplained}, and detailed in Line 2 of Alg.~\ref{algo:fast}.

Based on these sets, the most repeated element is selected, and the sets are split into two branches of a tree, (i)~without the most repeated element on the left and (ii)~with the most repeated element on the right, as shown in the function \textit{Binary\_Decision\_Tree} in Line 6 of Alg.~\ref{algo:fast}. The sets on the right are assumed to remove the most frequently occurring element as the common term. The branching on either side of the tree continues till there are no more repeated elements in the sets. In the aforementioned example, $D$ is the most repeated element. The only set without $D$, $\{A,B\}$, is written on the left branch. The other sets with $D$ are written on the right branch, assuming that $D$ is taken out as the common term, as $\{A,E,F\}$, $\{C,E,F\}$, and $\{C,B\}$ in the second step in Fig.~\ref{fig:fastexampleexplained}. This procedure of splitting corresponds to the \textit{Split} function in Line 18 of Alg.~\ref{algo:fast}.

The set on the left $\{A.B\}$ cannot be further split. However, the sets on the right can be further split with $C$ as the most repeated element, giving the third step in Fig.~\ref{fig:fastexampleexplained} based on the \textit{Split} function in Line 18 of Alg.~\ref{algo:fast}. No further branching is possible, and hence the obtained branches are leaf nodes. After the recursive splitting operations, the branches must be simplified before combining. For example, the leaf node $\{E,F\}$, and $\{B\}$ must be multiplied based on the \textit{EvaluateSelf} and \textit{Multiply} functions in Lines 57 and 66 of Alg.~\ref{algo:fast}, respectively. The two sets are multiplied and checked for absorption to remove any subsets. In this case, they are evaluated to be $\{BE, BF\}$. 
After the leaf nodes are evaluated, they must be combined using the \textit{Combine} function in Line 40 of Alg.~\ref{algo:fast}. For example, the branch with $\{A, E, F\}$ is combined with the aforementioned evaluation of the leaf node. The combination operation computes the union of each pair of elements from the two sets and checks for absorption, removing any resulting subsets if present. The combination operation is recursively repeated to identify the \acp{MCS} for the given source-destination pair. For the example topology in Fig.~\ref{fig:topo} and the source-destination pair $S-T$, the \acp{MCS} are $\{(A,D), (A,C), (B,D), (B,E), (B,F)\}$.

\subsection{Differentiating \fast~from Boole-Shannon expansion~\cite{rushdi2021derivation}}
Though our \fast~and the Boole-Shannon expansion used in~\cite{rushdi2021derivation} employ a recursive decomposition strategy, they differ fundamentally. The Boole-Shannon expansion is applied directly on the Boolean representation of the system's success function (\acp{MPS} in this work), manipulating logical expressions to derive \acp{MCS} as the prime implicants of the complemented function. On the other hand, our \fast~operates strictly on a set-theoretic representation of \acp{MPS} and never constructs an explicit Boolean formula. \fast's decision tree recursively partitions the sets into subsets that either contain or do not contain an element. 

For large topologies, handling very long Boolean expressions, expansions, and absorptions is memory and time-consuming. However, \fast's set operations are faster and more memory-efficient. This property is also clear from the results discussed in Sec.~\ref{chap:results}.

\subsection{Workflow}
\label{sec:workflow}
\begin{figure}
    \centering
    \includegraphics[width=1\linewidth]{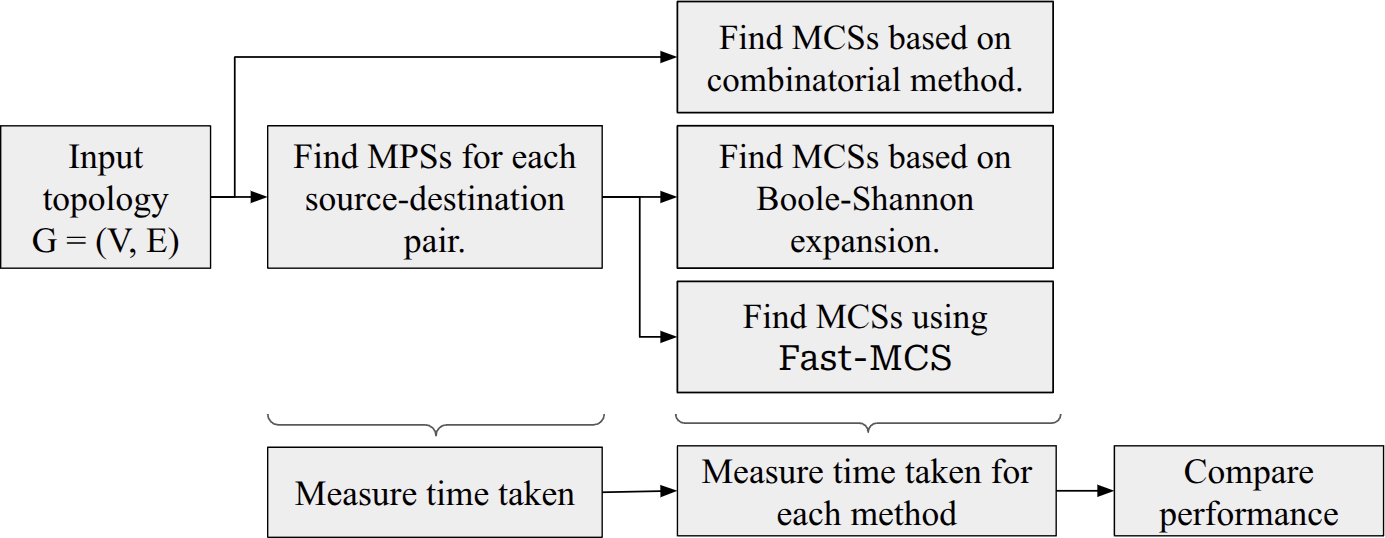}
    \caption{Workflow adopted in this work.}
    \label{fig:workflow}
\end{figure}

Fig.~\ref{fig:workflow} shows the workflow adopted in this work. Given an input topology, first, \acp{MPS} are evaluated for each source-destination pair. Then, the approaches to find \acp{MCS}- combinatorial method, Boole-Shannon expansion, and our \fast~are evaluated. The times required to find \acp{MPS} and \acp{MCS} using the different approaches are measured and then compared. 
\begin{table*}[]
\centering
\caption{Topologies and their sources, evaluated in this study. $|V|$ and $|E|$ denote the number of nodes and edges in topology $G$.}
\mini
\label{table:topos}
\begin{tabular}{|c|c|c|ccccccccc}
\hline
\textbf{$G$}              & \textbf{$|V|$} & \textbf{$|E|$} & \multicolumn{1}{c|}{\textbf{$G$}}           & \multicolumn{1}{c|}{\textbf{$|V|$}} & \multicolumn{1}{c|}{\textbf{$|E|$}} & \multicolumn{1}{c|}{\textbf{$G$}}            & \multicolumn{1}{c|}{\textbf{$|V|$}} & \multicolumn{1}{c|}{\textbf{$|E|$}} & \multicolumn{1}{c|}{\textbf{$G$}}           & \multicolumn{1}{c|}{\textbf{$|V|$}} & \multicolumn{1}{c|}{\textbf{$|E|$}} \\ \hline
Abilene \cite{zoo}        & 11             & 14             & \multicolumn{1}{c|}{Spain \cite{sai}}       & \multicolumn{1}{c|}{17}             & \multicolumn{1}{c|}{28}             & \multicolumn{1}{c|}{USA\_26 \cite{sai}}      & \multicolumn{1}{c|}{26}             & \multicolumn{1}{c|}{42}             & \multicolumn{1}{c|}{India35 \cite{sai}}     & \multicolumn{1}{c|}{35}             & \multicolumn{1}{c|}{80}             \\ \hline
Polska \cite{sndlib}      & 12             & 18             & \multicolumn{1}{c|}{Austria\_24 \cite{sai}} & \multicolumn{1}{c|}{24}             & \multicolumn{1}{c|}{50}             & \multicolumn{1}{c|}{Norway \cite{sai}}       & \multicolumn{1}{c|}{27}             & \multicolumn{1}{c|}{51}             & \multicolumn{1}{c|}{jonas-us-ca \cite{sai}} & \multicolumn{1}{c|}{39}             & \multicolumn{1}{c|}{61}             \\ \hline
HiberniaUK \cite{zoo}     & 13             & 13             & \multicolumn{1}{c|}{Sweden \cite{sai}}      & \multicolumn{1}{c|}{25}             & \multicolumn{1}{c|}{29}             & \multicolumn{1}{c|}{Nobel\_EU \cite{sndlib}} & \multicolumn{1}{c|}{28}             & \multicolumn{1}{c|}{41}             & \multicolumn{1}{c|}{pioro40 \cite{sai}}     & \multicolumn{1}{c|}{40}             & \multicolumn{1}{c|}{89}             \\ \hline
Germany\_17 \cite{sndlib} & 17             & 26             &                                             &                                     &                                     &                                              &                                     &                                     &                                             &                                     &                                     \\ \cline{1-3}
\end{tabular}
\squeezeup
\end{table*}

\section{Results}
\label{chap:results}

We evaluated \fast~on an AMD Ryzen 5 3600 hexacore processor with 24 GB RAM running WSL2 \cite{WSL2-Linux-Kernel} on 13 topologies shown in Table~\ref{table:topos}. For each topology, each possible source-destination pair is considered. The workflow for \fast~corresponds to that discussed in Fig.~\ref{fig:workflow}.

\subsection{\fast~vs. Combinatorial approach}
\begin{figure}
    \centering
    \includegraphics[width=1\linewidth]{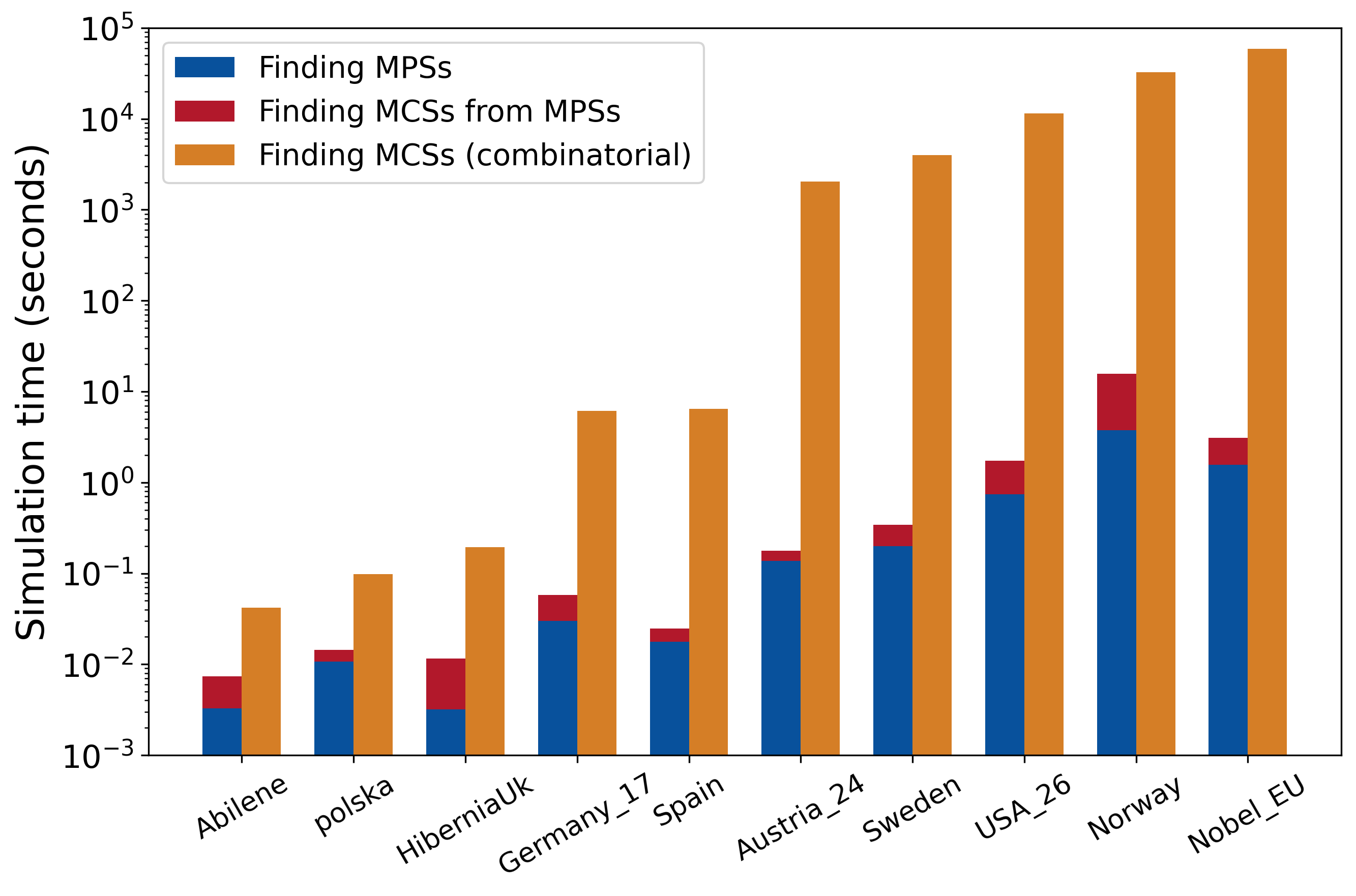}
    \caption{Comparing the runtimes for \fast~and the combinatorial approach~\cite{mine_pyrbdplusplus, alghanim} (Sec.~\ref{sec:combi}). The blue and red bars correspond to finding \acp{MPS} and \acp{MCS} from \acp{MPS}, respectively. Combined, the two bars together represent the total runtime for \fast. The orange bar corresponds to the combinatorial method. \fast~outperforms the combinatorial method in every topology.}
    \label{fig:fastmcs vs combinatorial}
    \squeezeup
\end{figure}

Fig.~\ref{fig:fastmcs vs combinatorial} shows \fast's performance compared against the traditional combinatorial approach~\cite{mine_pyrbdplusplus,  alghanim} as explained in Section~\ref{sec:combi}. The Y axis shows the simulation time in seconds on a logarithmic scale for the corresponding topology in the X axis. \fast~consists of two major steps- finding \acp{MPS} and then finding \acp{MCS} based on the \acp{MPS}. The former is represented by the blue bars, while the latter is represented by the red bars stacked over the blue bars. The red and blue bars together show the total time taken for our \fast~tool to evaluate \acp{MCS}. On the other hand, the orange bars show the time taken for the combinatorial approach. For smaller topologies, \fast~performs around a 100$\times$ faster, while larger topologies run over a 1000$\times$ faster. Therefore, Fig.~\ref{fig:fastmcs vs combinatorial} shows \fast's evident superior performance. The combinatorial approach could only be evaluated for 10 topologies owing to time constraints. 
\begin{figure}
    \centering
    \includegraphics[width=1\linewidth]{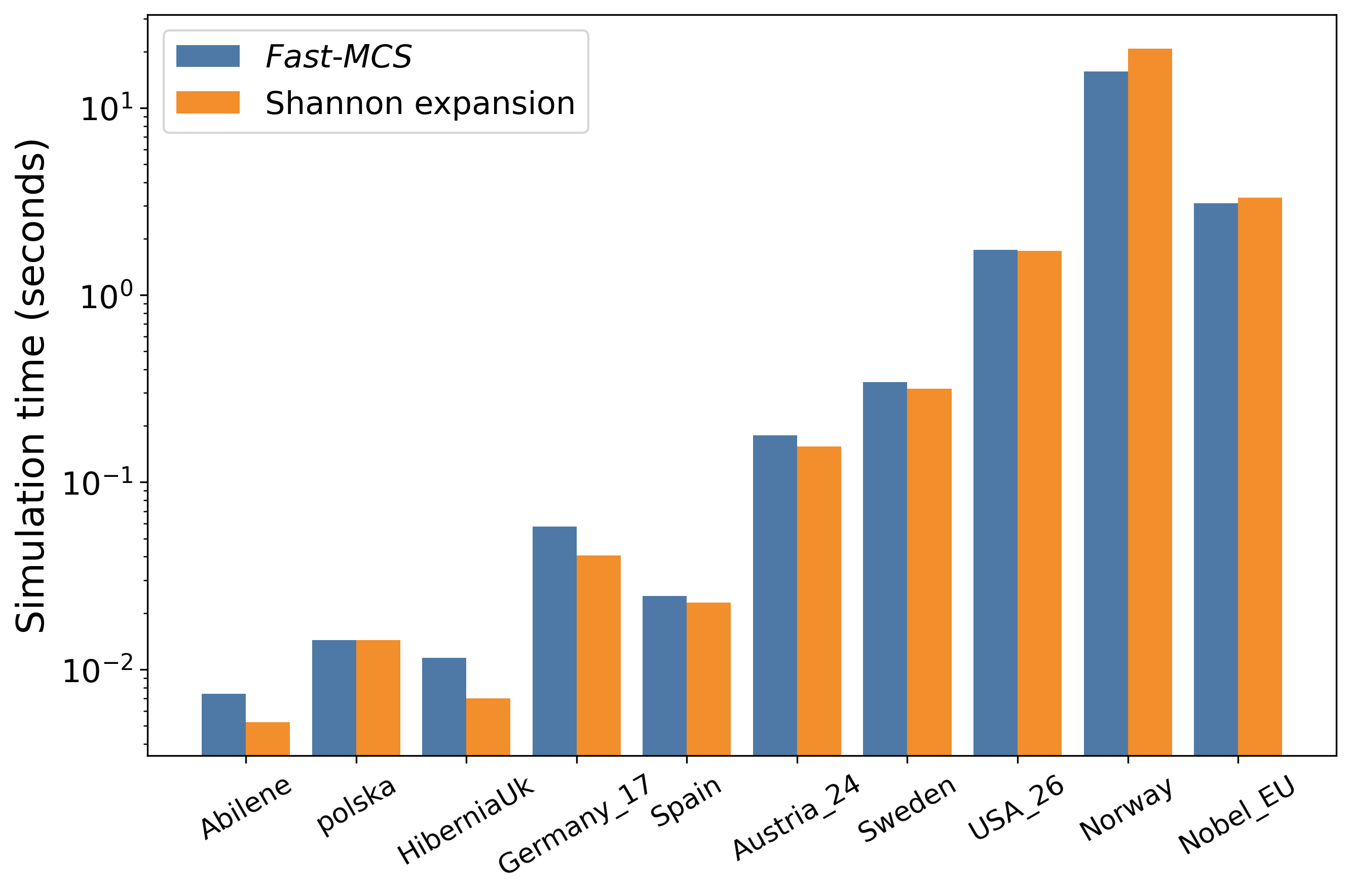}
    \caption{Comparing the runtimes for \fast~and the Boole-Shannon expansion approach~\cite{rushdi2021derivation} (Sec.~\ref{sec:shannon}) in the blue and orange bars, respectively. Boole-Shannon expansion approach offers slightly better performance than \fast~for small topologies.}
    \label{fig:fastmcs vs shannon - small}
    \squeezeup
\end{figure}
\begin{figure}
    \centering
    \includegraphics[width=1\linewidth]{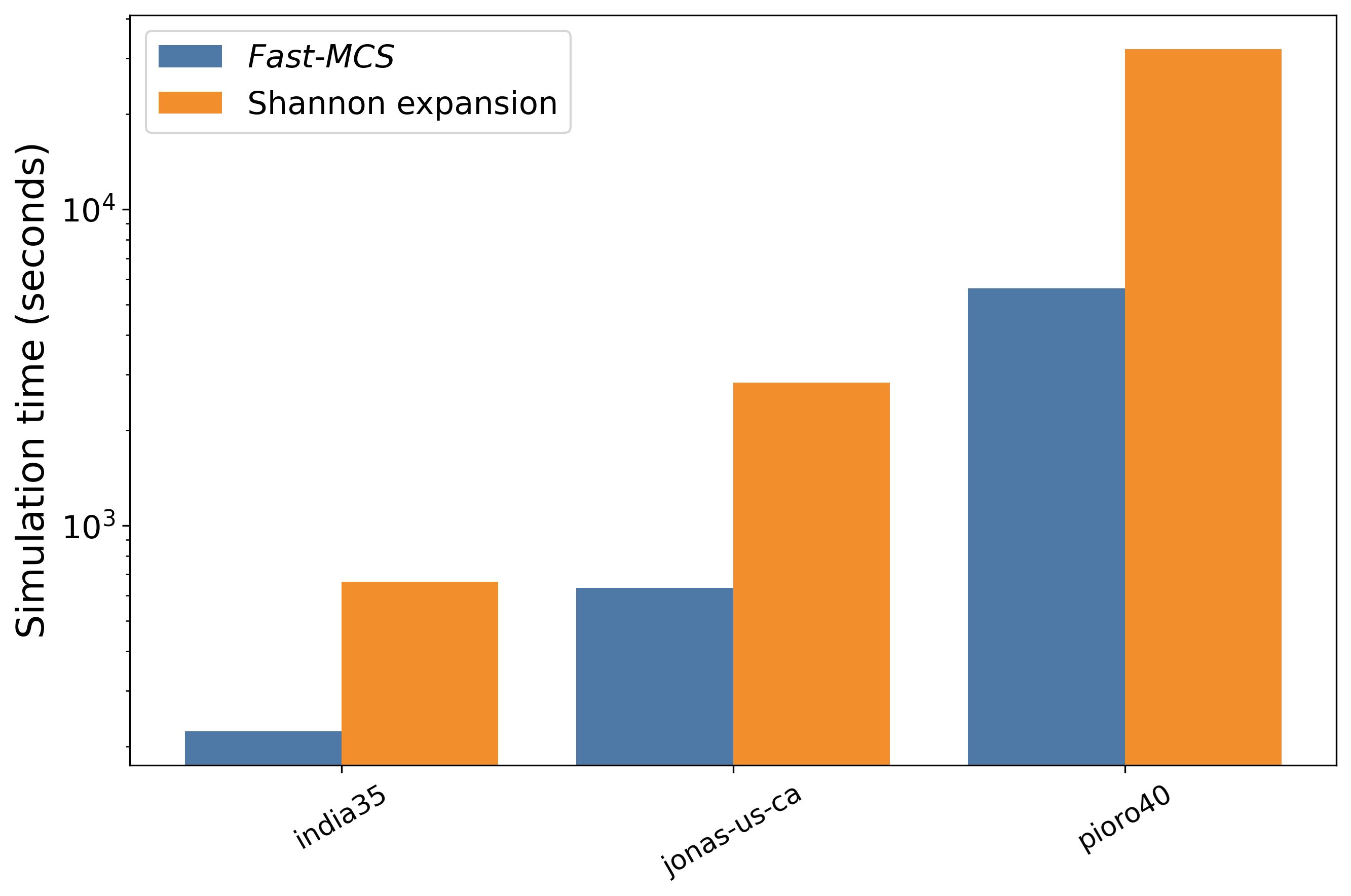}
    \caption{Comparing the runtimes for \fast~and the Boole-Shannon expansion approach~\cite{rushdi2021derivation} (Sec.~\ref{sec:shannon}) in the blue and orange bars, respectively. \fast~provides significantly faster computation than Boole-Shannon expansion approach.}
    \label{fig:fastmcs vs shannon - large}
    \squeezeup
\end{figure}

\subsection{\fast~vs.Boole-Shannon expansion}
Fig.~\ref{fig:fastmcs vs shannon - small} shows \fast's performance compared against the Shannon expansion method proposed in~\cite{rushdi2021derivation}. The Y axis shows the simulation time in seconds on a logarithmic scale for the corresponding topology in the X axis. The blue and orange bars correspond to the runtimes of \fast~and Shannon expansion techniques, respectively. Shannon expansion~\cite{rushdi2021derivation} outperforms \fast~in smaller topologies by less than 100 milliseconds. However, for larger topologies, \fast~performs better by a few seconds. To further investigate this behavior, we tested both techniques on larger topologies as seen in Fig.~\ref{fig:fastmcs vs shannon - large}. Here, \fast~outperforms Shannon expansion by at least 80\%. The runtime saved by using \fast~in larger topologies far outweighs the runtime lost by using \fast~in smaller topologies. Therefore, Fig.~\ref{fig:fastmcs vs shannon - large} shows that \fast~is scalable and efficient in large topologies. Figs.~\ref{fig:fastmcs vs combinatorial},~\ref{fig:fastmcs vs shannon - small} and~\ref{fig:fastmcs vs shannon - large} are plotted separately because the Y axis varies significantly for each. Plotting them together loses critical information portrayed in Fig.~\ref{fig:fastmcs vs shannon - large}. 

\section{Conclusions and Outlook}
\label{chap:conclusion}

\acfp{MCS} are crucial tools in evaluating a network's resilience.
In this work, we present our scalable, open-source tool, \fast, for efficiently obtaining \acp{MCS}. We compared \fast's performance against other methods, such as the combinatorial method and Shannon expansion, and show that \fast~is faster than the other methods. Though the examples considered in this manuscript are communication network topologies, \fast~is applicable to any system that can be represented as a graph. Currently, components are considered binary-valued, i.e., either working or unavailable. In the future, we intend to evaluate \acp{MCS} as a time-varying function of components' availabilities.


\bibliographystyle{IEEEtran}

\begin{thebibliography}{10}
\providecommand{\url}[1]{#1}
\csname url@samestyle\endcsname
\providecommand{\newblock}{\relax}
\providecommand{\bibinfo}[2]{#2}
\providecommand{\BIBentrySTDinterwordspacing}{\spaceskip=0pt\relax}
\providecommand{\BIBentryALTinterwordstretchfactor}{4}
\providecommand{\BIBentryALTinterwordspacing}{\spaceskip=\fontdimen2\font plus
\BIBentryALTinterwordstretchfactor\fontdimen3\font minus \fontdimen4\font\relax}
\providecommand{\BIBforeignlanguage}[2]{{%
\expandafter\ifx\csname l@#1\endcsname\relax
\typeout{** WARNING: IEEEtran.bst: No hyphenation pattern has been}%
\typeout{** loaded for the language `#1'. Using the pattern for}%
\typeout{** the default language instead.}%
\else
\language=\csname l@#1\endcsname
\fi
#2}}
\providecommand{\BIBdecl}{\relax}
\BIBdecl
\renewcommand{\BIBentryALTinterwordstretchfactor}{4}

\bibitem{mcs}
A.~Rauzy, ``Mathematical foundations of minimal cutsets,'' \emph{{IEEE} Trans. Rel.}, vol.~50, no.~4, pp. 389--396, 2001.

\bibitem{sndlib}
S.~Orlowski \emph{et~al.}, ``{SNDlib} 1.0-survivable network design library,'' vol.~55, no.~3.\hskip 1em plus 0.5em minus 0.4em\relax Wiley Online Library, 2010, pp. 276--286.

\bibitem{ft_mcs}
J.~Vatn, ``Finding minimal cut sets in a fault tree,'' \emph{Rel. Eng. \& Syst. Saf.}, vol.~36, no.~1, pp. 59--62, 1992.

\bibitem{faulttree_analaysis_1}
E.~Ruijters and M.~Stoelinga, ``Fault tree analysis: A survey of the state-of-the-art in modeling, analysis and tools,'' \emph{Comput. Sci. Rev.}, vol.~15, pp. 29--62, 2015.

\bibitem{fiabilipy}
\BIBentryALTinterwordspacing
V.~Lecrubier, ``Fiabilipy,'' GitHub repository. [Online]. Available: \url{https://github.com/crubier/Fiabilipy}
\BIBentrySTDinterwordspacing

\bibitem{ft2}
Z.~Tang and J.~B. Dugan, ``Minimal cut set/sequence generation for dynamic fault trees,'' in \emph{Annu. Symp. Rel. and Maintainability (RAMS)}.\hskip 1em plus 0.5em minus 0.4em\relax IEEE, 2004, pp. 207--213.

\bibitem{relyence}
\BIBentryALTinterwordspacing
{Relyence}, ``{Relyence RBD}.'' [Online]. Available: \url{https://relyence.com/products/rbd/}
\BIBentrySTDinterwordspacing

\bibitem{mine_pyrbdplusplus}
S.~Janardhanan \emph{et~al.}, ``{PyRBD++: An Open-Source Fast Reliability Block Diagram Evaluation Tool},'' \emph{15th Int. Workshop on Resilient Netw. Des. and Model. (RNDM)}, 2025.

\bibitem{mine_pyrbd3}
------, ``Leveraging minimal path sets for reliability block diagram evaluation,'' in \emph{IEEE Future Netw. World Forum (FNWF)}, 2025.

\bibitem{7}
R.~Patil, ``An overview of fault tree analysis ({FTA}) method for reliability analysis,'' \emph{J. of Eng. Res. and Stud.}, vol.~4, pp. 6--8, 03 2013.

\bibitem{mine6_ICTON-demo}
S.~Janardhanan \emph{et~al.}, ``Improving network sovereignty - a minimal cut set approach,'' in \emph{24th Int. Conf. on Transparent Opt. Netw. (ICTON)}, 2024, pp. 1--4.

\bibitem{mine_commag}
\BIBentryALTinterwordspacing
------, ``How to build a sovereign network?-a proposal to measure network sovereignty,'' \emph{arXiv preprint arXiv:2510.23510}, 2025. [Online]. Available: \url{https://arxiv.org/abs/2510.23510}
\BIBentrySTDinterwordspacing

\bibitem{mine_GUI}
------, ``Interactive demonstration of an open-source dependability suite for communication networks,'' in \emph{25th Int. Conf. on Transparent Opt. Netw. (ICTON)}, 2025, pp. 1--4.

\bibitem{mine_PyRobust}
------, ``{PyRobust: An Open-Source Robustness Surface Generation Tool},'' in \emph{25th Int. Conf. on Transparent Opt. Netw. (ICTON)}, 2025, pp. 1--4.

\bibitem{yeh}
\BIBentryALTinterwordspacing
W.-C. Yeh, ``A simple algorithm to search for all mcs in networks,'' \emph{Eur. J. of Oper. Res.}, vol. 174, no.~3, pp. 1694--1705, 2006. [Online]. Available: \url{https://www.sciencedirect.com/science/article/pii/S0377221705002511}
\BIBentrySTDinterwordspacing

\bibitem{mine_pyrbd}
S.~Janardhanan \emph{et~al.}, ``{PyRBD: An Open-Source Reliability Block Diagram Evaluation Tool},'' in \emph{IEEE Int. Workshop Tech. Committee on Commun. Qual. and Rel. (CQR)}, 2024, pp. 19--24.

\bibitem{alghanim}
\BIBentryALTinterwordspacing
A.~M. Al-Ghanim, ``A heuristic technique for generating minimal path and cutsets of a general network,'' \emph{Comput. \& Ind. Eng.}, vol.~36, no.~1, pp. 45--55, 1999. [Online]. Available: \url{https://www.sciencedirect.com/science/article/pii/S0360835298001119}
\BIBentrySTDinterwordspacing

\bibitem{kvassay_analysis_2016}
\BIBentryALTinterwordspacing
M.~Kvassay \emph{et~al.}, ``\BIBforeignlanguage{en}{Analysis of minimal cut and path sets based on direct partial {Boolean} derivatives},'' \emph{\BIBforeignlanguage{en}{Proc. of the Inst. of Mech. Engineers, Part O: J. of Risk and Rel.}}, vol. 230, no.~2, pp. 147--161, Apr. 2016. [Online]. Available: \url{https://journals.sagepub.com/doi/10.1177/1748006X15598722}
\BIBentrySTDinterwordspacing

\bibitem{43}
\BIBentryALTinterwordspacing
T.~Eiter \emph{et~al.}, ``Computational aspects of monotone dualization: A brief survey,'' \emph{Discrete Appl. Math.}, vol. 156, no.~11, pp. 2035--2049, 2008, in Memory of Leonid Khachiyan (1952 - 2005 ). [Online]. Available: \url{https://www.sciencedirect.com/science/article/pii/S0166218X07001278}
\BIBentrySTDinterwordspacing

\bibitem{46}
\BIBentryALTinterwordspacing
L.~Khachiyan \emph{et~al.}, ``An efficient implementation of a quasi-polynomial algorithm for generating hypergraph transversals and its application in joint generation,'' \emph{Discrete Appl. Math.}, vol. 154, no.~16, pp. 2350--2372, 2006, discrete Algorithms and Optimization, in Honor of Professor Toshihide Ibaraki at His Retirement from Kyoto University. [Online]. Available: \url{https://www.sciencedirect.com/science/article/pii/S0166218X06001910}
\BIBentrySTDinterwordspacing

\bibitem{heidtmann}
K.~Heidtmann, ``Inverting paths \& cuts of 2-state systems,'' \emph{{IEEE} Trans. Rel.}, vol. R-32, no.~5, pp. 469--474, 1983.

\bibitem{rai}
S.~Rai and K.~Aggarwal, ``On complementation of pathsets and cutsets,'' \emph{{IEEE} Trans. Rel.}, vol. R-29, no.~2, pp. 139--140, 1980.

\bibitem{locks}
M.~O. Locks, ``Inverting and minimalizing path sets and cut sets,'' \emph{{IEEE} Trans. Rel.}, vol. R-27, no.~2, pp. 107--109, 1978.

\bibitem{prasad}
\BIBentryALTinterwordspacing
V.~Prasad \emph{et~al.}, ``Generation of vertex and edge cutsets,'' \emph{Microelectron. Rel.}, vol.~32, no.~9, pp. 1291--1310, 1992. [Online]. Available: \url{https://www.sciencedirect.com/science/article/pii/0026271492906533}
\BIBentrySTDinterwordspacing

\bibitem{lamalem}
Y.~Lamalem \emph{et~al.}, ``New and fast algorithm to minimal cutsets enumeration based on necessary minimal paths,'' in \emph{5th Int. Symp. on Innov. in Inf. and Commun. Technol. (ISIICT)}, 2018, pp. 1--5.

\bibitem{rushdi2021derivation}
A.~Rushdi and M.~H. Amashah, ``Derivation of minimal cutsets from minimal pathsets for a multi-state system and utilization of both sets in checking reliability expressions,'' \emph{J. of Eng. Res. and Rep.}, vol. 20(8), pp. 22--33, 2021.

\bibitem{mine_git_fastmcs}
S.~Janardhanan and Y.~Chen, ``\texttt{Fast-MCS},'' \url{www.github.com/shakthij98/fastmcs}.

\bibitem{zoo}
S.~Knight \emph{et~al.}, ``The internet topology zoo,'' \emph{{IEEE} J. Sel. Areas Commun.}, vol.~29, no.~9, pp. 1765 --1775, Oct. 2011.

\bibitem{sai}
S.~K. Patri, ``Github, \textit{Physical Network Information},'' \url{www.github.com/SaiPatri/PhyNWInfo}.

\bibitem{WSL2-Linux-Kernel}
\BIBentryALTinterwordspacing
Microsoft, ``{WSL2-Linux-Kernel},'' 2025. [Online]. Available: \url{\url{https://github.com/microsoft/WSL2-Linux-Kernel}}
\BIBentrySTDinterwordspacing

\end{thebibliography}

\begin{acronym}
   \acro{RBD}[RBD]{Reliability Block Diagram}
   \acro{MCS}[MCS]{Minimal Cut Set}
   \acro{MPS}[MPS]{Minimal Path Set}
   \acro{CDM}[CDM]{Conditional Decomposition Method}
   \acro{DFS}[DFS]{Depth First Search}
   \acro{RC}[RC]{Relative Complement}
\end{acronym}

\end{document}